\documentclass[11pt,superscriptaddress,aps,prd,preprint]{revtex4}

\everymath{\displaystyle}
\usepackage{graphicx}

\usepackage{amsmath,amssymb,mathrsfs}

\newcommand{\bea}{\begin{eqnarray}}
\newcommand{\eea}{\end{eqnarray}}

\begin{document}

\title{On the Gravitational Energy of Axial Perturbations in Regular Black Holes}

\author{S. C. Ulhoa}\email[]{sc.ulhoa@gmail.com}
\affiliation{Instituto de F\'isica, Universidade de Bras\'ilia, 70910-900, Bras\'ilia, DF, Brazil} \affiliation{Canadian Quantum Research Center,\\ 
106-460 Doyle Ave, Kelowna, British Columbia V1Y 0C2 Canada} 

\author{F. L. Carneiro}\email[]{fernandolessa45@gmail.com}
\affiliation{Universidade Federal do Norte do Tocantins,
Centro de Ci\^encias Integradas,
77824-838 Aragua\'ina, TO, Brazil}

\author{B. C. C. Carneiro}\email[]{bcccarneiro@gmail.com}
\affiliation{Instituto Federal do Tocantins,  77760-000, Colinas do Tocantins, TO, Brazil}

\begin{abstract}

The article deals with the gravitational energy associated with axial perturbations of regular black holes. We review the stability of the geometry under odd-parity perturbations and the corresponding quasinormal modes, previously obtained for this class of spacetimes. The perturbative functions describing the metric fluctuations are reconstructed from the master equation. To evaluate the energy content of these perturbations, we employ the Teleparallel Equivalent of General Relativity (TEGR), which provides a well-defined expression for gravitational energy. The gravitational energy is computed up to second order in the perturbation parameter and expressed in terms of the quasinormal mode functions. Our results establish a direct connection between the dynamical response of regular black holes and the energy carried by their gravitational perturbations.

\end{abstract}

\date{\today}

\maketitle

\section{Introduction}

It is well known that the equations of General Relativity admit solutions that contain physical singularities, regions where geometric invariants diverge and the very causal structure of spacetime ceases to be well defined. The first exact solution ever found, due to Karl Schwarzschild~\cite{Schwarzschild1916}, already exhibited this type of behavior inside the black hole. This is a physical singularity in the sense that no coordinate transformation can remove it. The presence of singularities was early recognized as an indication of the limitations of the classical theory, having led Albert Einstein himself to question whether such solutions could represent physically realistic situations~\cite{Einstein1939, Weinstein2023}. From a conceptual point of view, singularities are problematic because they break the causal relation between events, preventing any consistent physical description of the region where the theory ceases to be valid. This aversion to singularities also motivated the formulation of the cosmic censorship conjecture~\cite{Penrose1969, Penrose1999}, proposed by the group led by Roger Penrose, according to which nature would not produce naked singularities.

Equally important as the problem of singularities is the issue of the stability of the solutions of Einstein's equations. The Schwarzschild solution was soon recognized as describing a new kind of physical object, the black hole. An object of this nature, formed through gravitational collapse and possessing a gravitational field strong enough to prevent even light from escaping, cannot simply become unstable when it captures matter from its surroundings. This accretion process must evolve toward a stable final state after some time; otherwise, such objects could hardly possess physical reality. In this context, Tullio Regge and John Archibald Wheeler analyzed the conditions under which the Schwarzschild solution remains stable under gravitational perturbations~\cite{ReggeWheeler1957}. They showed that axial perturbations can be described by a Schr\"odinger-type master equation with an effective potential determined by the background geometry. Later, Frank Zerilli obtained the corresponding master equation for polar perturbations~\cite{Zerilli1970}. It was then realized that the time dependence of the solutions to these equations involves complex frequencies, so that the oscillations are damped over time. This behavior defines the so-called quasinormal modes~\cite{Vishveshwara1970}, which describe how black holes dynamically respond to perturbations and carry the imprint of the perturbed geometry. In principle, any solution of Einstein's equations can be subjected to this type of stability analysis.

The problem of singularities in the solutions of General Relativity is conceptually profound. From a philosophical perspective, it may be interpreted as an indication of the incompleteness of the theory, in the sense that a unified field theory would not exhibit such features. However, this does not imply that the presence of singularities is an unavoidable consequence of the geometric formulation of gravitation. In fact, regular solutions describing black-hole–like structures are already known. The first regular black hole solution was obtained by James Bardeen through the coupling of a magnetic monopole to the gravitational field equations~\cite{bardeen}. The resulting spacetime is free of a central singularity, while still possessing an event horizon. Subsequently, other regular solutions were proposed~\cite{hayward}, based not on exotic objects such as magnetic monopoles, but on extensions of classical electrodynamics to nonlinear theories. Regular black holes thus become natural candidates for the interpretation of astrophysical observations that point to the existence of extremely compact massive objects.

Regular black holes may also play an important role in cosmological scenarios. Certain classes of regular black holes with de Sitter cores and their remnants have been proposed as viable candidates for heavy dark matter, providing possible observational signatures associated with primordial inhomogeneities and the evolution of the early Universe~\cite{Dymnikova2015}. Furthermore, primordial black holes may naturally arise during first-order cosmological phase transitions, where collisions of vacuum bubbles can generate false-vacuum configurations whose subsequent collapse leads to black-hole formation with high probability~\cite{Konoplich1999,Khlopov2000}. These possibilities reinforce the physical relevance of investigating the dynamical properties, stability, and energy content of regular black-hole geometries.

The possibility that physically realistic black holes may be regular motivates the investigation of the stability of such objects under gravitational perturbations. At first sight, one would not expect a magnetic monopole to serve as the source of a stable gravitational configuration. However, somewhat surprisingly, regular black holes admit a master equation for axial perturbations and therefore allow the existence of quasinormal modes of vibration. The first result in this direction was obtained in~\cite{Ulhoa2014}, where a Regge--Wheeler--type equation is derived, albeit with an angular coupling different from that appearing in the Schwarzschild black hole case. Despite the success in showing that a regular black hole can represent a stable solution, several open issues remain to be explored. The most evident one is the absence of a complete analysis for all polarizations: a polar perturbation leading to a Zerilli--type master equation has not yet been established for this spacetime. The second open issue is precisely the one we address in the present work, namely, the determination of the gravitational energy associated with axial perturbations.

When discussing gravitational energy, it is important to recall that this is one of the oldest problems in General Relativity. The first attempts to define a gravitational energy--momentum tensor can be traced back to the structural analyses of the field equations carried out by Albert Einstein himself. However, a fundamental obstacle soon emerged: the resulting quantities took the form of pseudotensors. Over time, arguments accumulated suggesting that gravitational energy might not be localizable, casting doubt on the very physical meaning of an energy associated with the gravitational field. Even the Hamiltonian formulation of the theory was unable to establish an unambiguous expression for gravitational energy~\cite{adm}. On the other hand, it is now understood that this difficulty is closely related to the purely metric formulation in which General Relativity was originally developed. The Teleparallel Equivalent of General Relativity (TEGR) provides a dynamically equivalent description of gravitation that does not suffer from the same limitations in dealing with gravitational energy. Within this framework, there exists a well-defined expression for the energy of the gravitational field, whose consistency has been explored and confirmed over the years~\cite{maluf}. In this article, we employ this formalism to obtain the energy associated with axial perturbations of regular black holes.

This article is organized as follows. In Section~\ref{sec.2}, we review the spacetime of a regular black hole and derive the master equation governing axial gravitational perturbations, obtaining the corresponding quasinormal modes. In Section~\ref{sec.3}, we present a brief overview of the Teleparallel Equivalent of General Relativity (TEGR), emphasizing the definition of gravitational energy within this framework. In Section~\ref{sec.4}, we construct the tetrad field associated with the perturbed spacetime and compute the gravitational energy up to second order in the perturbation parameter, expressing the result in terms of the axial perturbation functions obtained earlier. Finally, in Section~\ref{sec.5}, we discuss the physical implications of the results and present our conclusions. Throughout this work, we adopt natural units, such that G = c = 1.

\section{Quasinormal Modes of a Regular Black Hole}\label{sec.2}

In this section we briefly review the main results obtained in Ref.~\cite{Ulhoa2014}. A regular black hole can be described by the following line element

\begin{equation}
ds^2 = -f(r)\, dt^2 + \frac{1}{f(r)}\, dr^2 + r^2\, d\theta^2 + r^2 \sin^2\theta\, d\phi^2 ,
\label{eq:metric}
\end{equation}
where
\begin{equation*}
f(r) = 1 - \frac{2M r^2}{(r^2 + \alpha^2)^{3/2}} .
\end{equation*}
This line element corresponds to the Bardeen class of regular black holes, where the parameter $\alpha$ controls the deviation from the Schwarzschild geometry. This spacetime is regular at $r=0$, since all curvature invariants remain finite at the origin. Nevertheless, it may still possess an event horizon, defined by the condition $f(r)=0$.

A solution of Einstein's equations is said to be stable under gravitational perturbations if such perturbations decay with time. In general, the perturbed metric can be written as
\begin{equation*}
g_{\mu\nu} = \bar{g}_{\mu\nu} + \epsilon\,h_{\mu\nu} ,
\end{equation*}
where $\bar{g}_{\mu\nu}$ denotes the background metric, in this case the regular black hole spacetime, and $h_{\mu\nu}$ represents the gravitational perturbations. These perturbations can be decomposed into two sectors, axial (odd--parity) and polar (even--parity).

The axial perturbations are defined by the off-diagonal components
\begin{equation*}
\delta g_{0\phi} = \epsilon\, h_0(r,t)\, h(\theta), 
\qquad 
\delta g_{r\phi} = \epsilon\, h_1(r,t)\, h(\theta) .
\end{equation*}
We assume a harmonic time dependence,
\begin{equation}
h_0(r,t) = \tilde{h}_0(r)\, e^{-i\omega t}, 
\qquad 
h_1(r,t) = \tilde{h}_1(r)\, e^{-i\omega t} ,
\label{eq:harmonic}
\end{equation}
and
\begin{equation*}
h(\theta) = P_\ell(\cos\theta) .
\end{equation*}
The parameter $\epsilon$ is introduced as a bookkeeping quantity controlling the perturbative order, and the metric is considered up to first order in $\epsilon$. It is worth noting that the angular dependence of the perturbation does not follow the original Regge--Wheeler ansatz. Instead, an a priori unknown angular function was introduced and subsequently determined from Einstein's equations, leading to Legendre polynomials rather than their derivatives.

Substituting the perturbations into Einstein's equations and performing the standard separation of variables, the system reduces to a single second-order differential equation for a master variable. It reads
\begin{equation*}
\frac{d^2 \phi}{dx^2} + \left( \omega^2 - V(x) \right) \phi = 0 ,
\end{equation*}
where
\begin{equation*}
\phi(r) = \frac{f(r)\,\tilde{h}_1(r)}{r} ,
\end{equation*}
and the effective potential is given by
\begin{equation*}
V(r) = f(r) \left[ \frac{\ell(\ell+1)}{r^2} + \frac{2\bigl(f(r)-1\bigr)}{r^2} + \frac{f'(r)}{r} + f''(r) + 16\pi\, L(r) \right] ,
\end{equation*}
with
\begin{equation*} 
L(r) = \frac{3M \alpha^5 r^3}{(r^2 + \alpha^2)^{5/2}} .
\end{equation*}
The master equation is written in terms of derivatives with respect to the tortoise coordinate, which is defined by
\begin{equation*}
\frac{dx}{dr} = \frac{1}{f(r)} .
\end{equation*} 
This choice removes the coordinate singularity at the horizon and allows the imposition of boundary conditions corresponding to purely ingoing waves at the event horizon and purely outgoing waves at spatial infinity.

The third-order WKB approximation can be employed to determine the quasinormal frequencies, yielding
\begin{equation}
\omega^2_{n,\ell} = V_0 + \left( -2 V''_0 \right)^{1/2} \Lambda 
- i \left( n + \frac{1}{2} \right) \left( -2 V''_0 \right)^{1/2} \left( 1 + \Omega \right),
\label{eq:WKBfreq}
\end{equation}
where
\begin{equation*}
\Lambda = \frac{1}{\left( -2 V''_0 \right)^{1/2}} \left[
\frac{1}{8} \frac{V^{(4)}_0}{V''_0} \left( \frac{1}{4} + \beta^2 \right)
- \frac{1}{288} \left( \frac{V'''_0}{V''_0} \right)^2 (7 + 60 \beta^2)
\right],
\end{equation*}
and
\begin{align*}
\Omega &= \frac{1}{-2 V''_0} \Bigg[
\frac{5}{6912} \left( \frac{V'''_0}{V''_0} \right)^4 (77 + 188 \beta^2)
- \frac{1}{384} \frac{(V'''_0)^2 V^{(4)}_0}{(V''_0)^3} (51 + 100 \beta^2) \nonumber\\
&\quad + \frac{1}{2304} \left( \frac{V^{(4)}_0}{V''_0} \right)^2 (67 + 68 \beta^2)
+ \frac{1}{288} \frac{V'''_0 V^{(5)}_0}{(V''_0)^2} (19 + 28 \beta^2)
- \frac{1}{288} \frac{V^{(6)}_0}{V''_0} (5 + 4 \beta^2)
\Bigg].
\end{align*}
Here $V_0$ and its derivatives are evaluated at the maximum of the potential barrier. We note that $\beta = n + \frac{1}{2}$, where $n$ is an integer. As a consequence, the quasinormal frequencies are discrete and complex, the imaginary part of the frequency determines the damping rate of the oscillations, ensuring the stability of the perturbations.

Within the same WKB approximation, the wave function can be written in the asymptotic form
\begin{equation*}
\phi_{\rm out}(r) = \mathcal{A}(r)\, e^{i S(r)}, 
\qquad 
\phi_{\rm in}(r) = \mathcal{A}(r)\, e^{-i S(r)} .
\end{equation*}
Here the phase function is given by
\begin{equation*}
S(r) = \int_{r_0}^{r} \frac{p(r')}{f(r')}\, dr' ,
\end{equation*}
where
\begin{equation*}
p(r) = \sqrt{\omega^2 - V(r)} ,
\end{equation*}
and the WKB amplitude reads
\begin{equation*}
\mathcal{A}(r) = \frac{1}{\sqrt{p(r)}} .
\end{equation*}
With these expressions, the perturbation functions can be reconstructed from the solution of the master equation as
\begin{equation}
\tilde{h}_1(r) = \frac{r}{f(r)}\, \phi(r) ,
\end{equation}
and
\begin{equation}
\tilde{h}_0(r) = -\frac{f(r)}{i\omega}\, \frac{d}{dr}(r\phi).
\end{equation}
We thus obtain both the quasinormal frequencies and the explicit radial functions $h_0$ and $h_1$ describing the gravitational perturbations. These quantities will be used in Section~\ref{sec.4} to compute the gravitational energy of the perturbations.

\section{Teleparallelism Equivalent to General Relativity (TEGR)} \label{sec.3}

The formulation of TEGR adopted here follows the approach presented in Ref.~\cite{maluf}. From a dynamical point of view, TEGR is equivalent to General Relativity. This means that the field equations are the same and, consequently, both theories share the same set of solutions. However, TEGR provides a well-defined alternative description of gravitational energy, which we now proceed to outline.

The temporal component of the tetrad field is always tangent to the worldline of a given observer. This property, when generalized, allows the tetrads to be interpreted as a reference frame adapted to a specific observer. In fact, there exists a direct relation between the metric tensor and the tetrads, given by
\begin{equation*}
g_{\mu\nu} = e^{a}{}_{\mu}\, e_{a\nu} .
\end{equation*}
It is important to note that the metric tensor has ten independent components due to its symmetry, whereas the tetrad field possesses sixteen components. The six additional components correspond precisely to the choice of reference frame. It is worth emphasizing that the metric tensor alone does not determine the underlying geometric structure. In particular, an affine connection must be specified in order to define covariant differentiation and parallel transport. The Christoffel symbols are uniquely determined by the metric only after imposing metric compatibility and vanishing torsion. In the teleparallel description one adopts instead the curvature-free Weitzenb\"ock connection, while keeping the same metric structure.

Consider a spacetime endowed with the connection
\begin{equation}
\Gamma_{\mu\lambda\nu} = e^{a}\,_{\mu}\, \partial_{\lambda} e_{a\nu} ,
\end{equation}
whose torsion tensor is
\begin{equation}
T^{a}\,_{\lambda\nu} = \partial_{\lambda} e^{a}\,_{\nu} - \partial_{\nu} e^{a}\,_{\lambda} .
\label{4}
\end{equation}
This spacetime is known as a Weitzenb\"ock space. It exhibits a complementary structure with respect to Riemannian geometry. In Riemannian geometry, the connection is given by the Christoffel symbols, the curvature is nonzero, and torsion vanishes. In contrast, in Weitzenb\"ock geometry the connection defined above (often referred to as the Weitzenb\"ock connection) has vanishing curvature and nonzero torsion. This contrast naturally raises the question of whether a relation exists between these two connections. Indeed, they are connected through the identity
\begin{equation*}
\Gamma_{\mu\lambda\nu} = {}^{0}\Gamma_{\mu\lambda\nu} + K_{\mu\lambda\nu} ,
\label{2}
\end{equation*}
where ${}^{0}\Gamma_{\mu\lambda\nu}$ denotes the Christoffel symbols and
\begin{equation}
K_{\mu\lambda\nu} = \frac{1}{2}\left( T_{\lambda\mu\nu} + T_{\nu\lambda\mu} + T_{\mu\lambda\nu} \right)
\label{3}
\end{equation}
is the contortion tensor of the Weitzenb\"ock spacetime.

The identity relating the Weitzenb\"ock and Riemannian connections also establishes a relation between the curvature scalar constructed from the Christoffel symbols and the torsion tensor. This relation can be written as
\begin{equation*}
eR(e)\equiv -e\left(\frac{1}{4}T^{abc}T_{abc}+\frac{1}{2}T^{abc}T_{bac}-T^aT_a\right)+2\partial_\mu(eT^\mu)\,,
\label{eq5}
\end{equation*}
where $T^\mu=T^{\nu\mu}\,_{\nu}$ and $e$ denotes the determinant of the tetrad field, that is, $e=\sqrt{-g}$. It is important to note that the left-hand side is precisely the Hilbert--Einstein Lagrangian density. This motivates the choice of the right-hand side, up to a total divergence, as the Lagrangian density of TEGR. Since a total divergence does not affect the field equations, we adopt the Lagrangian density
\begin{eqnarray}
\mathfrak{L}(e_{a\mu})&=& -\kappa\,e\left(\frac{1}{4}T^{abc}T_{abc}+
\frac{1}{2} T^{abc}T_{bac} -T^aT_a\right) -\mathfrak{L}_M\nonumber \\
&\equiv&-\kappa\,e \Sigma^{abc}T_{abc} -\mathfrak{L}_M\;,
\label{6}
\end{eqnarray}
where
\begin{equation}
\Sigma^{abc}=\frac{1}{4} (T^{abc}+T^{bac}-T^{cab}) +\frac{1}{2}(
\eta^{ac}T^b-\eta^{ab}T^c)\;,
\label{7}
\end{equation}
and $\mathfrak{L}_M$ represents the Lagrangian density of the matter fields. In natural units, the gravitational coupling constant is $\kappa=\frac{1}{16\pi}$.

Varying the Lagrangian density with respect to the tetrad field, which is the dynamical variable of the theory, yields
\begin{equation}
\partial_\nu\left(e\Sigma^{a\lambda\nu}\right)=\frac{1}{4\kappa}
e\, e^a\,_\mu( t^{\lambda \mu} + T^{\lambda \mu})\;,
\label{10}
\end{equation}
where
\begin{equation}
t^{\lambda \mu}=\kappa\left[4\,\Sigma^{bc\lambda}T_{bc}\,^\mu- g^{\lambda
\mu}\, \Sigma^{abc}T_{abc}\right]
\label{11}
\end{equation}
is interpreted as the gravitational energy--momentum tensor. It can be shown that the TEGR field equations are equivalent to Einstein’s equations. The interpretation of $t^{\lambda\mu}$ follows from the antisymmetry of $\Sigma^{a\lambda\nu}$ (often called the superpotential) in its last two indices, which implies the identity
\begin{equation*}
\partial_\lambda\partial_\nu\left(e\Sigma^{a\lambda\nu}\right)\equiv0\,.
\label{12}
\end{equation*}
This represents a conservation law for the right-hand side of Eq.~(\ref{10}). Consequently, a natural energy--momentum vector is defined as
\begin{equation}
P^a = \int_V d^3x \,e\,e^a\,_\mu(t^{0\mu}+ T^{0\mu})\,,
\label{14}
\end{equation}
or, using the field equations, in terms of a surface integral as
\begin{equation}
P^a =4\kappa\, \int_V d^3x \,\partial_\nu\left(e\,\Sigma^{a0\nu}\right)\,.
\label{14.1}
\end{equation}
This definition of the energy--momentum vector is fully covariant under Lorentz transformations (associated with Latin indices) and invariant under coordinate transformations (associated with Greek indices). Unlike pseudotensorial constructions, it is expressed entirely in terms of well-defined tensors and is not restricted to asymptotic limits, as in the ADM approach. The absence of pseudotensorial ambiguities is a direct consequence of the use of tetrad fields and torsion as fundamental geometric quantities. In the next section we will use this definition to compute the energy associated with the axial perturbations of a regular black hole.

\section{Energy of Axial Perturbations of a Regular Black Hole}\label{sec.4}

In order to compute the energy associated with axial perturbations, one must first fix a reference frame. Here we adopt the same tetrad field used in Ref.~\cite{MalufUlhoa2021}, where the energy of polar and axial perturbations was evaluated. The tetrad field adapted to stationary observers is given by
\begin{equation}
e_{a\mu}
=
\begin{pmatrix}
 -A & 0 & 0 & -D \\
 0 & B \sin\theta \cos\phi & r \cos\theta \cos\phi &
     - E\, r \sin\theta \sin\phi + F \sin\theta \cos\phi \\
 0 & B \sin\theta \sin\phi & r \cos\theta \sin\phi &
     \phantom{-}E\, r \sin\theta \cos\phi + F \sin\theta \sin\phi \\
 0 & B \cos\theta & - r \sin\theta & F \cos\theta
\end{pmatrix},
\label{eq:tetrada}
\end{equation}
where
\begin{align*}
A &= \sqrt{-g_{00}}, \\
B &= \sqrt{g_{11}}, \\
D &= - \frac{g_{03}}{\sqrt{-g_{00}}}, \\
E^2 &= 1 + \frac{g_{11} g_{03}^2 + g_{00} g_{13}^2}{r^2 \sin^2\theta}, \\
F &= g_{13}\sqrt{g_{11}} .
\end{align*}
With this choice, the observer four-velocity is
$U^\mu = e_{(0)}{}^\mu = \left(\frac{1}{A},0,0,0\right)\,.$

We now compute the gravitational energy associated with the axial perturbations using the definition of the energy--momentum vector introduced in the previous section. Since the metric is written as a perturbative expansion around the regular black hole background, the tetrad field, torsion tensor and superpotential also admit expansions in powers of the bookkeeping parameter $\epsilon$. Consequently, the energy $P^{(0)}$ naturally splits into a background contribution and perturbative corrections. The background spacetime corresponds to $\epsilon=0$ and describes the unperturbed regular black hole. We are interested only in the energy carried by the gravitational perturbations. Therefore, we define the variation of the gravitational energy as the leading nonvanishing perturbative contribution, namely the term of second order in $\epsilon$ after subtracting the background value. The first-order contribution in $\epsilon$ vanishes identically, so that the leading perturbative energy arises at order $\epsilon^2$, as expected for a quadratic energy functional. This quantity is written as
\begin{equation*}
\delta \mathcal{E}(r)
\equiv
\frac{1}{\epsilon^{2}}
\left[
P^{(0)}(r)\big|_{O(\epsilon^2)} - P^{(0)}(r)\big|_{\epsilon=0}
\right].
\end{equation*}
After substituting the perturbed tetrad into the expression (\ref{14.1}) and keeping terms up to order $\epsilon^2$, the gravitational energy variation takes the form
\begin{equation}\label{deltaE}
\delta \mathcal{E}(r)
=
\int_{0}^{\pi} d\theta \;
\frac{r^{-1}}{32}\,
\left(\frac{1}{f(r)}\right)^{5/2}
\, h(\theta)^2
\;\mathcal{I}(r,\theta),
\end{equation}
where the integrand is given by
\[
\begin{aligned}
\mathcal{I}(r,\theta)
=&\;
f(r)^3\, h_1(r,t)^2 \csc\theta
\left[
-2\cos^2\theta
\left(
-1+\frac{1}{\sqrt{f(r)}}
\right)
-\frac{r\left(-2-\sqrt{f(r)}+f(r)\right)f'(r)}{f(r)}
\right]
\\[6pt]
&+
h_0(r,t)^2 \csc\theta
\left[
\frac{2\cos^2\theta}{\sqrt{f(r)}}
-4r f'(r)
- r\sqrt{f(r)} f'(r)
+ f(r)\left(-2\cos^2\theta + r f'(r)\right)
\right]
\\[6pt]
&-
4 r\, f(r)\,
h_0(r,t)\,
\sin^2\theta\,
\partial_t h_1(r,t)
\;.
\end{aligned}
\]
It should be emphasized that in Eq. (\ref{deltaE}) the quantity r is not treated as a dynamical coordinate, but as a parameter specifying the radius of the closed two--surface arising from the surface integral in Eq. (\ref{14.1}). In order to extract the physical content of this result, we evaluate $\delta\mathcal{E}(r)$ numerically and analyze its radial behavior for representative quasinormal modes. Because the analytical form of the integrand is rather involved, a direct interpretation is more transparent through graphical analysis. The following plots illustrate how the gravitational energy associated with the axial perturbations is distributed in the exterior region of the regular black hole. We also compare these profiles with the corresponding behavior found in the Schwarzschild case discussed in Ref.~\cite{MalufUlhoa2021}, highlighting how the regularity of the spacetime modifies the energy distribution of the perturbations. Since quasinormal modes have complex frequencies, the energy density is also complex. In the plots we show the real part, which captures the oscillatory spatial structure of the energy distribution.

\begin{figure}[ht]
\centering
\includegraphics[width=0.85\textwidth]{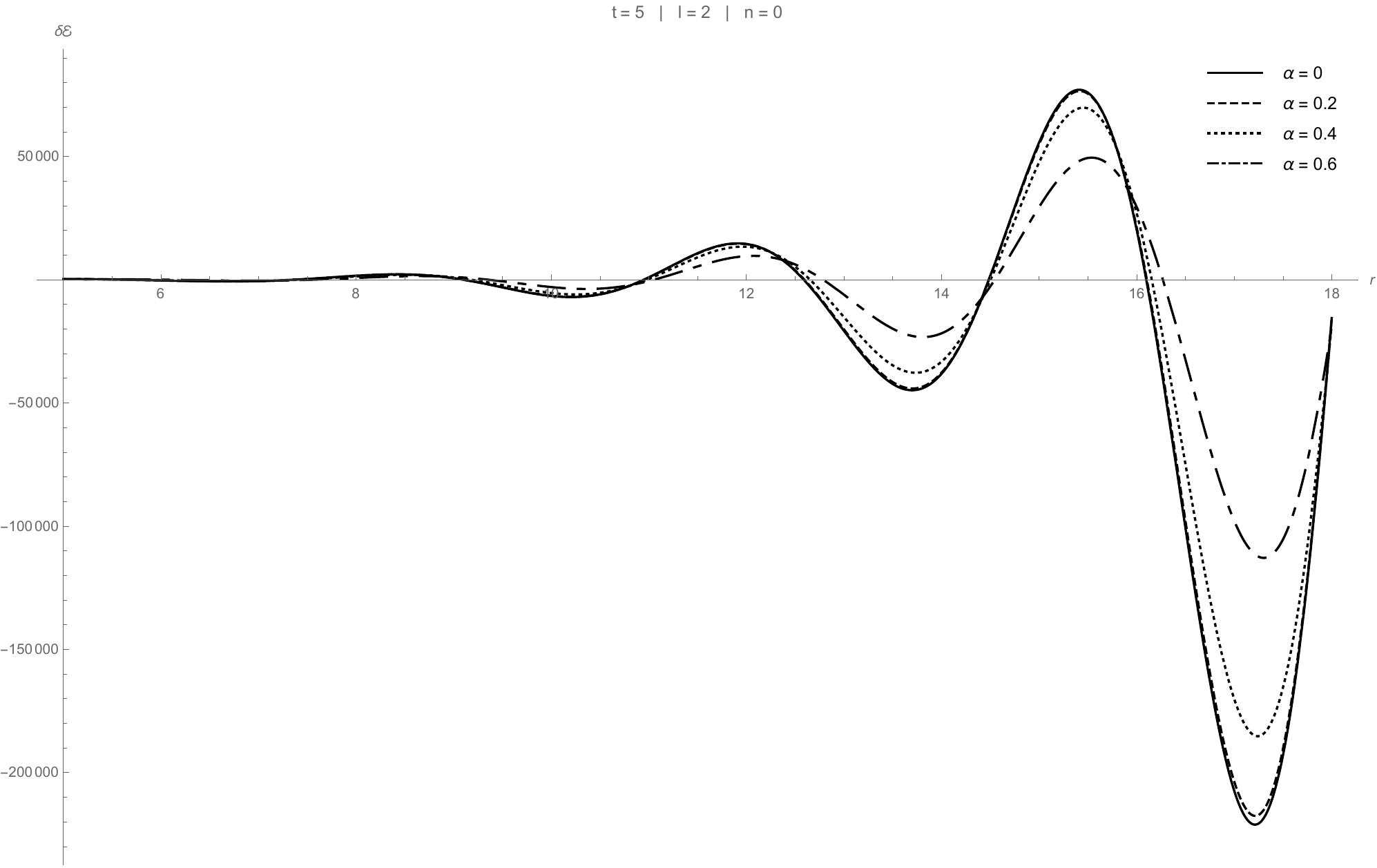}
\caption{
Radial profile of the real part of the gravitational energy variation $\mathrm{Re}\,\delta\mathcal{E}(r)$ for the fundamental axial quasinormal mode with $\ell=2$ and $n=0$, evaluated at fixed time $t=5$. The mass is set to $M=1$ and different curves correspond to distinct values of the regularization parameter $\alpha$.  
}
\label{fig:energy_alpha}
\end{figure}

\begin{figure}[ht]
\centering
\includegraphics[width=0.85\textwidth]{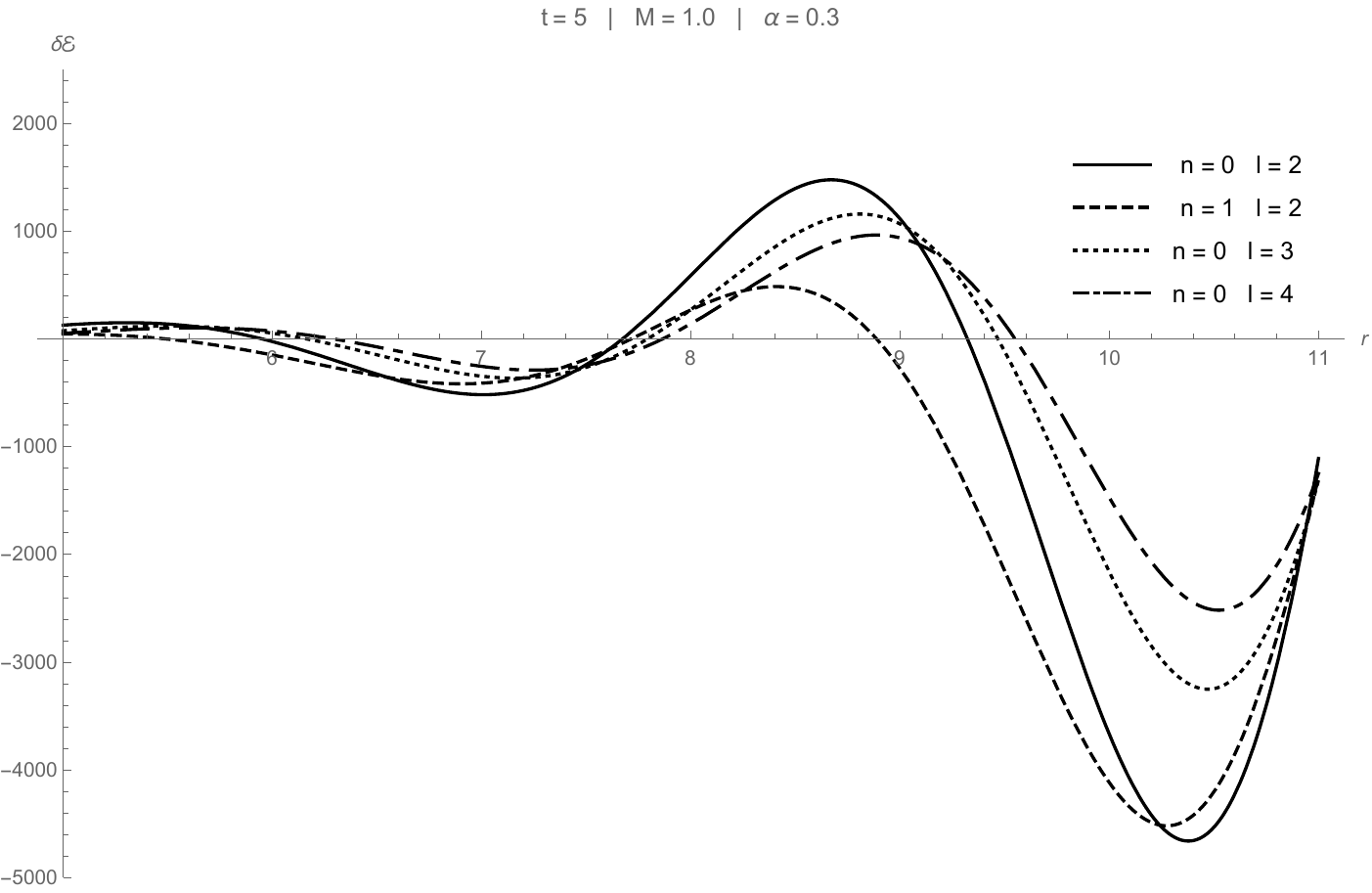}
\caption{
Radial profile of the real part of the gravitational energy variation $\mathrm{Re}\,\delta\mathcal{E}(r)$ for several axial quasinormal modes, evaluated at fixed time $t=5$ with $M=1$ and $\alpha=0.3$. The different curves correspond to modes with $(\ell,n)=(2,0)$, $(2,1)$, $(3,0)$, and $(4,0)$, whose frequencies determine the spatial oscillation pattern. 
}
\label{fig:energy_omega}
\end{figure}

Figures~\ref{fig:energy_alpha} and~\ref{fig:energy_omega} display the radial distribution of the gravitational energy associated with axial perturbations for representative quasinormal modes. In Fig.~\ref{fig:energy_alpha}, where the angular momentum number and overtone are fixed, varying the regularization parameter $\alpha$ primarily modifies the overall amplitude of the energy density while preserving the oscillatory structure. This indicates that the regular core mainly softens the strength of the gravitational response without significantly altering the spatial pattern of the perturbations. In contrast, Fig.~\ref{fig:energy_omega} shows that, for fixed $\alpha$, changes in the quasinormal frequency mainly affect the phase of the radial oscillations. Modes with larger real part of $\omega$ exhibit shorter spatial wavelengths, whereas the imaginary part governs the temporal damping and therefore has a weaker influence on the instantaneous spatial profile at fixed time. In both figures, the growth of the oscillations at large $r$ reflects the non-normalizable character of quasinormal modes, which behave as outgoing waves in the asymptotic region. 

\begin{figure}[ht]
\centering
\includegraphics[width=0.85\textwidth]{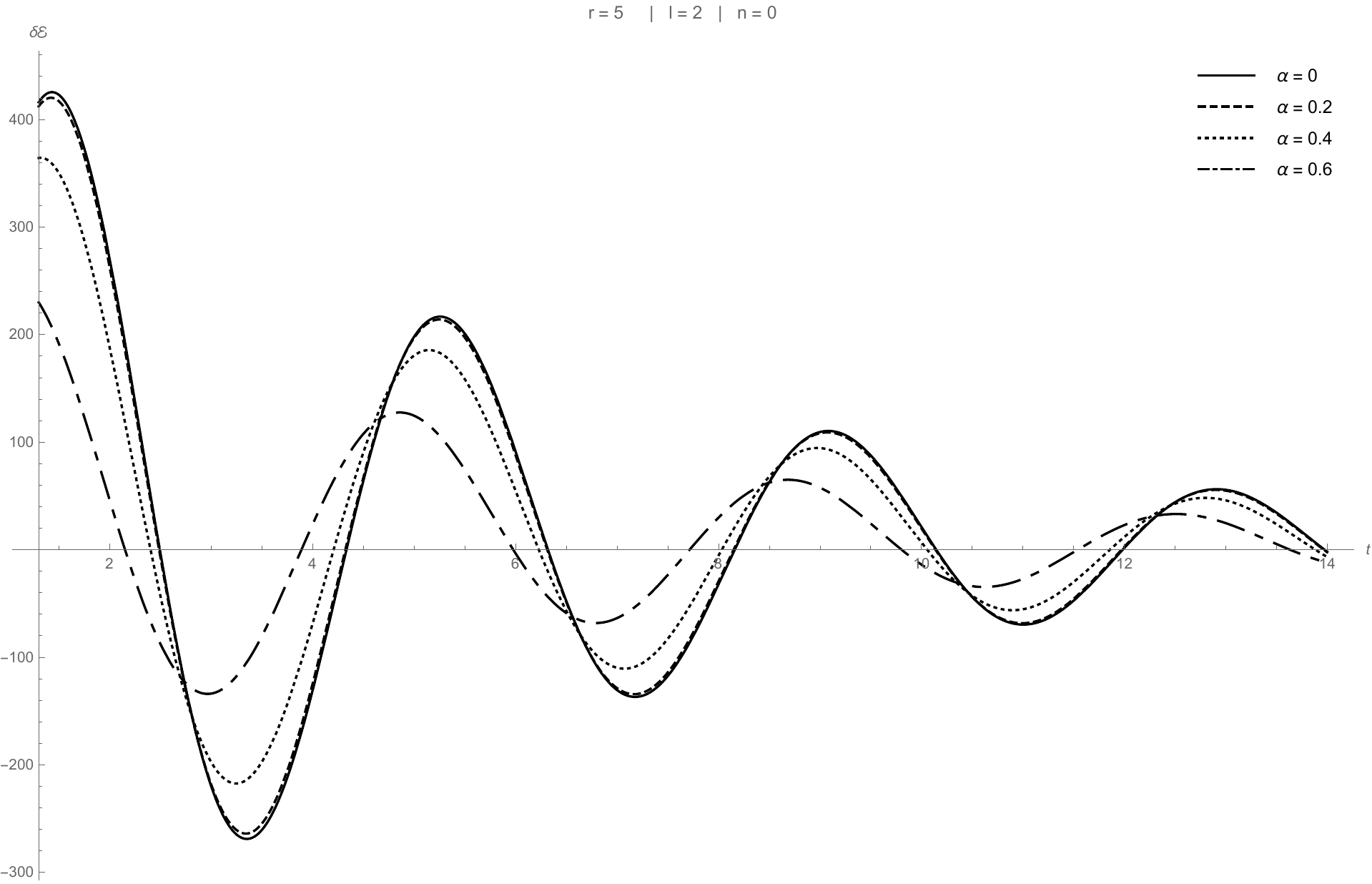}
\caption{
Temporal evolution of the real part of the gravitational energy variation $\mathrm{Re}\,\delta\mathcal{E}(t)$ evaluated at fixed radius $r=5$ for the fundamental axial quasinormal mode with $\ell=2$ and $n=0$. The mass is set to $M=1$, and the curves correspond to different values of the regularization parameter $\alpha$.
}
\label{fig:energy_time_alpha}
\end{figure}

\begin{figure}[ht]
\centering
\includegraphics[width=0.85\textwidth]{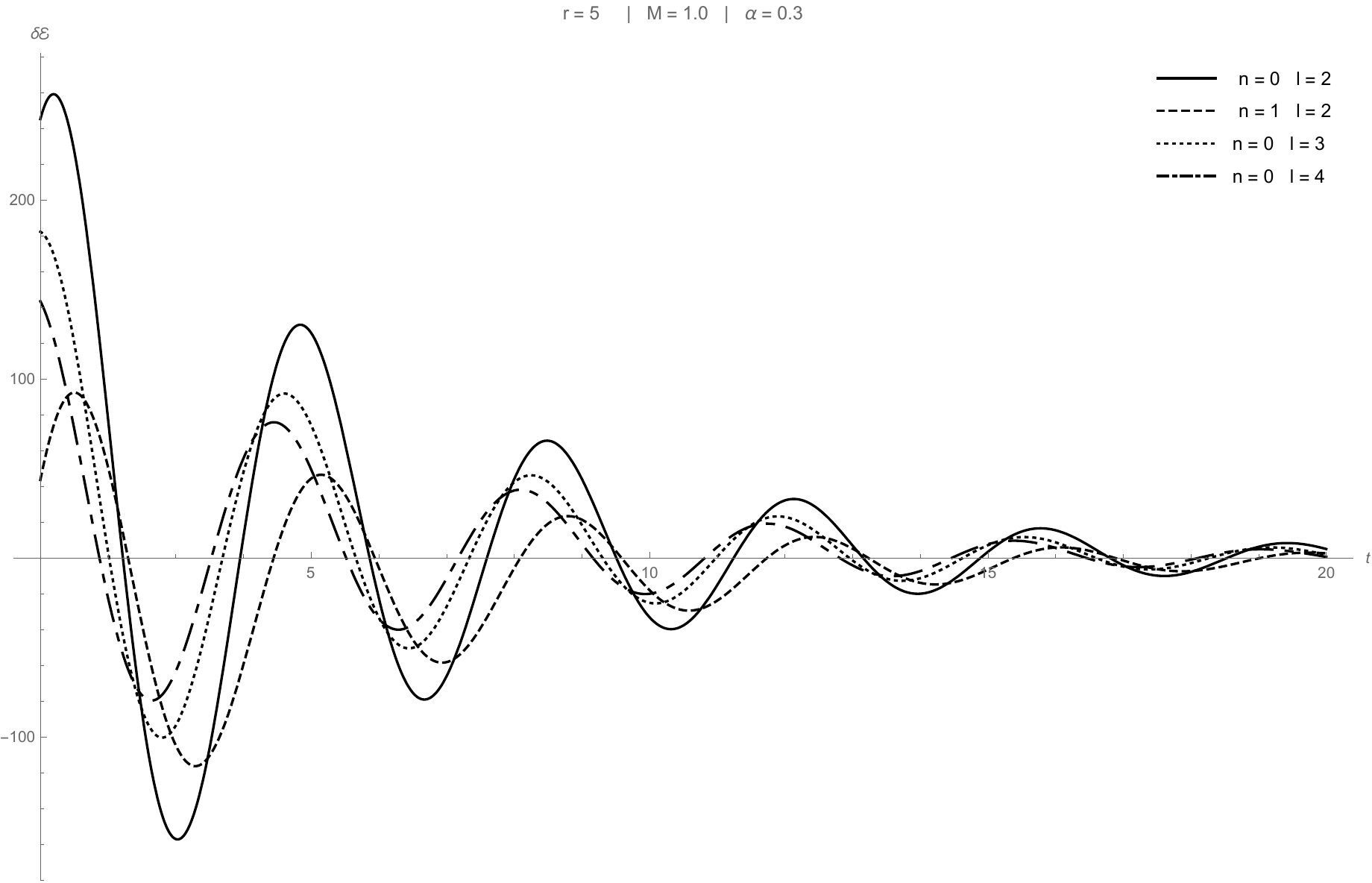}
\caption{
Temporal evolution of the real part of the gravitational energy variation $\mathrm{Re}\,\delta\mathcal{E}(t)$ evaluated at fixed radius $r=5$ with $M=1$ and $\alpha=0.3$. The curves correspond to different axial quasinormal modes with $(\ell,n)=(2,0)$, $(2,1)$, $(3,0)$, and $(4,0)$.
}
\label{fig:energy_time_omega}
\end{figure}

Having analyzed the spatial distribution of the gravitational energy, we now investigate its temporal evolution. This complementary perspective allows us to understand how the energy carried by the perturbations changes in time at a fixed radial position.
Figures~\ref{fig:energy_time_alpha} and~\ref{fig:energy_time_omega} show the temporal behavior of the gravitational energy associated with the axial perturbations at a fixed radial position. The energy exhibits the characteristic damped oscillatory pattern expected from quasinormal modes, reflecting the complex nature of the mode frequencies. In Fig.~\ref{fig:energy_time_alpha}, the oscillation frequency remains essentially unchanged as $\alpha$ varies, while the amplitude of the energy signal is affected by the regularization parameter. This indicates that the core regularization influences the intensity of the gravitational response without significantly modifying the oscillation timescale. Figure~\ref{fig:energy_time_omega} shows that different quasinormal modes produce distinct oscillation frequencies and damping rates in the energy signal. Modes with larger real part of the frequency oscillate more rapidly, while those with larger imaginary part decay more quickly in time. This behavior confirms that the temporal evolution of the gravitational energy directly encodes the quasinormal spectrum of the background geometry.

When $\alpha=0$, the background reduces to the Schwarzschild geometry, and both the radial and temporal energy profiles reproduce the qualitative behavior previously obtained in Ref.~\cite{MalufUlhoa2021}, up to an overall normalization factor. This provides an important consistency check of the present calculation.

\section{Conclusion} \label{sec.5}

In this work we have investigated the gravitational energy associated with axial perturbations of a regular black hole spacetime within the framework of the Teleparallel Equivalent of General Relativity (TEGR). The quasinormal spectrum employed here was not derived anew, but taken from our previous analysis of axial perturbations of regular black holes~\cite{Ulhoa2014}. Building upon those results, we reconstructed the perturbative metric functions and obtained an explicit algebraic expression for the gravitational energy up to second order in the perturbation parameter.

The resulting energy variation was analyzed both spatially and temporally. The radial profiles reveal how the regularization parameter $\alpha$ mainly affects the amplitude of the energy distribution, while variations in the quasinormal frequency predominantly modify the oscillatory structure. The temporal analysis shows the expected damped oscillatory behavior, directly reflecting the complex nature of the quasinormal frequencies. In all plots we considered low values of $\ell$ and overtone number $n$, which is consistent with the regime of validity of the WKB approximation used to obtain the quasinormal spectrum.

The present study provides a concrete link between the dynamical response of regular black holes and the gravitational energy carried by their perturbations. A natural extension of this work is the analysis of polar perturbations within the same framework. However, the existence and characterization of quasinormal modes for polar perturbations of regular black holes remain an open issue, representing an important gap in the current literature. Clarifying this point will be essential for a complete understanding of the gravitational energy associated with all perturbative sectors of regular black hole spacetimes.

\end{document}